\documentclass[prc,showpacs,amsmath,amssymb,preprint,superscriptaddress]{revtex4}
\usepackage{graphicx}

\newcommand{\nin}{Ni+Ni~}
\newcommand{\au}{Ni+Au~}
\newcommand{\so}{asy-soft~}
\newcommand{\st}{asy-stiff~}
\newcommand{\ex}{$E_{diss}$~}
\newcommand{\excm}{$E_{diss}/E_{c.m.}$~}

\begin{document}

\title
{Isospin diffusion in semi-peripheral $^{58}Ni$+$^{197}Au$ collisions 
at intermediate energies (II): Dynamical simulations}

\author{E.~Galichet}
\email[Corresponding author: ]{galichet@ipno.in2p3.fr}
\affiliation{Institut de Physique Nucl\'eaire, Universit\'e
Paris-Sud 11, CNRS/IN2P3, F-91406 Orsay Cedex, France}
\affiliation{Conservatoire National des Arts et M\'etiers , F-75141
Paris Cedex 03, France}
\affiliation{INFN Laboratori Nazionali del Sud, I-95123 Catania, Italy.}
\affiliation{INFN Sez. di Catania and Dipartimento di Fisica,
Universit\`a di Catania, Italy.}
\author{M.~Colonna}
\affiliation{INFN Laboratori Nazionali del Sud, I-95123 Catania, Italy.}
\author{B.~Borderie}
\author{M.~F.~Rivet}
\affiliation{Institut de Physique Nucl\'eaire, Universit\'e
Paris-Sud 11, CNRS/IN2P3, F-91406 Orsay Cedex, France}

\date{\today}

\begin{abstract}
We study isospin effects in semi-peripheral collisions above the Fermi 
energy by
considering the symmetric $^{58}Ni$+$^{58}Ni$ and the asymmetric
reactions $^{58}Ni$+$^{197}Au$
over the incident energy range 52-74~A~MeV. 
A microscopic transport model with two different parameterizations of the
symmetry energy term is used to investigate  the 
isotopic content of pre-equilibrium emission and the N/Z diffusion process.
Simulations are also compared to experimental data obtained with the INDRA
array and bring information on the degree of isospin equilibration
observed in Ni + Au collisions. A better overall agreement between data and
simulations is obtained when using a symmetry term which linearly increases with
nuclear density.
\end{abstract}


\maketitle

\section{Introduction}

Collisions between nuclei with different charge asymmetries may carry important information on
the structure of the nuclear equation of state (EOS) symmetry term in density
regions away from the normal value, that may be encountered along the reaction
path~\cite{BaoAnBook01,baranPR}.  For instance, the symmetry energy 
behaviour influences reaction processes, such as fragmentation, 
pre-equilibrium emission, N/Z equilibration between the two collisional 
partners~\cite{XuPRL85,Bao97,BaranNPA703,ShiPRC68,TsangPRL92,She04,Bao-An,Riz05,Fam06}.
Among the sensitive observables, in semi-peripheral collisions, one can look
at the isotopic content of light particle and IMF emission and at the asymmetry (N/Z) of the
reconstructed quasi-projectiles (QP) and quasi-targets (QT) 
\cite{TsangPRL92,She04,galichet}.
The degree of equilibration, that is related to the interplay between the
reaction time and the typical time for isospin transport,
can give information about important transport properties, such as
drift and diffusion coefficients, and their relation with the 
density dependence of the symmetry energy.   

In this paper we undertake this kind of investigations by studying 
isospin transport effects on the reaction dynamics in collisions with impact
parameter between 4 and 12 fm. Two systems, with the same 
projectile, $^{58}$Ni, and two different targets ($^{58}$Ni and $^{197}$Au), 
are considered at incident energies of 52 AMeV and 74 AMeV.
The N/Z ratio of the two composite systems is N/Z=1.07 for Ni+Ni 
and N/Z=1.38 for Ni+Au. 
The choice of the two systems and beam energies will allow us to
study isospin effects in different conditions of charge (and mass)
asymmetry and how they evolve as a function of the energy deposited 
into the system.   
In the symmetric Ni + Ni system  isospin effects are essentially due to the 
pre-equilibrium emission.
On the contrary, in the charge (and mass) asymmetric reactions, 
one can observe isospin transport between the two partners.
The dependence of these mechanisms on the symmetry energy behaviour is
discussed.\\ 

The paper is divided into three sections. In a first part (Section II), we describe the model used
and we present the results obtained, then we discuss the role of the isospin degree of freedom on the 
reaction dynamics and the comparison with some experimental data (Section III).
Conclusions are drawn in Section IV.

\section{Results of BNV code}

\subsection{Evolution in phase space}

\begin{figure}[htbp]
\resizebox{0.7\textwidth}{!}{%
\includegraphics{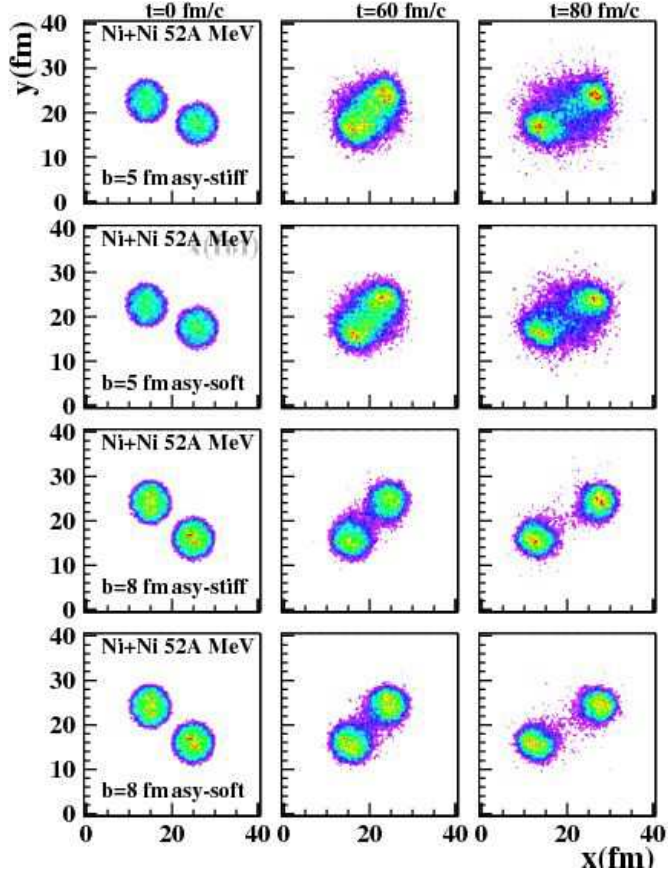}}
\caption{(color online) Density plots for \nin collisions at 52 AMeV}
\label{figure1}
\end{figure}

\begin{figure}[htbp]
\resizebox{0.7\textwidth}{!}{%
\includegraphics{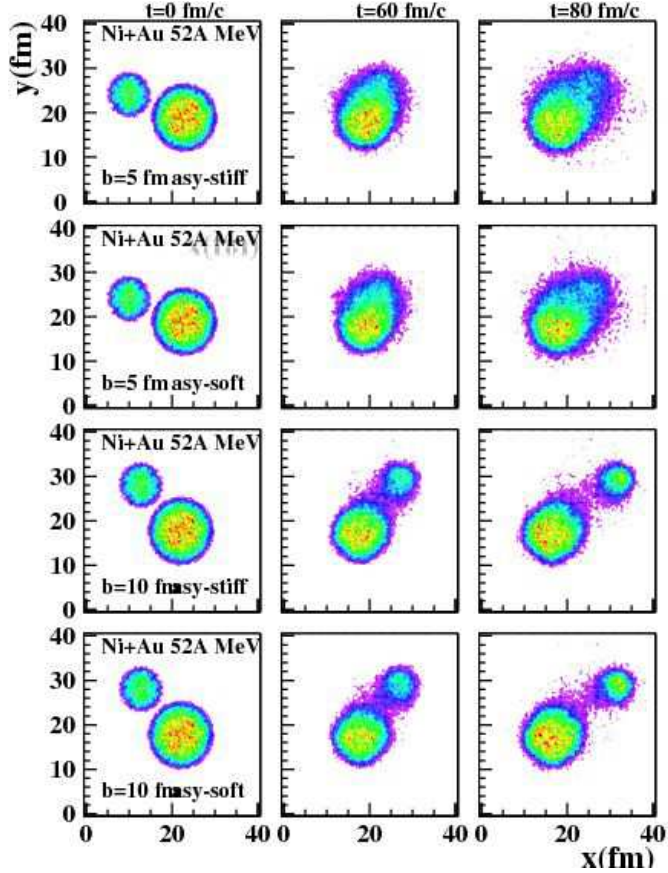}}
\caption{(color online) Density plots for \au collisions at 52 AMeV}
\label{figure12}
\end{figure}

We follow the reaction dynamics solving the BNV transport equation, 
that describes the evolution of the one-body distribution function 
according to the nuclear mean-field and including the effects of
two-body collisions \cite{UehlingPR43}.  
The test-particle prescription is adopted, using the TWINGO code
\cite{guarnera}.
The main ingredients that enter this equation are the nuclear matter 
compressibility, the symmetry energy term and its density dependence
and the nucleon-nucleon cross section.  
Here we will consider a compressibility modulus K =200 MeV and two different 
prescriptions for the behaviour of the symmetry energy,
in order to study the sensitivity of the results to the considered parameterization: an
``asy-stiff'' case for which the potential symmetry term
linearly increases with nuclear density 
($E_{sym}(\rho)=E_{sym}(\rho_0)(\rho/\rho_0)$), where $\rho_0$ is the
nuclear saturation density, and an ``asy-soft'' case
using the $SKM^*$ parameterization which exhibits a ($\rho/\rho_0)^{0.6}$
dependence (see~\cite{BaranNPA703} for more details). The free
nucleon-nucleon cross section with its angular, energy and isospin
dependence was used. For the two reactions, we have ran different impact 
parameters, from $b=4$ fm to $b=10$ fm for the \nin system
and from $b=4$ fm to $b=12$ fm for the \au system. For each impact parameter
 10 events were produced (one event represents already the mean trajectory of
the reaction), for the two cases of symmetry energy parameterization. 
In the following, except for figures~\ref{figure1} and~\ref{figure12},
the results shown will be averaged over the 10 events 
for each impact parameter. This reduces the fluctuations due to
the use of a finite number of test particles in the simulations.
In figures~\ref{figure1},~\ref{figure12} is displayed
the time evolution of density contours in the reaction plane,  for two  
impact parameters and the two parameterizations of the symmetry energy term, 
\so and \st{}.\\
In general the asy-stiff EOS can be linked to a 
more repulsive dynamics. Indeed, in this case, the system feels a 
stronger repulsion in the first stage of the collisions, due to the 
increased value of the symmetry energy above normal density. 
However, in proton-rich systems, the larger value of $E_{sym}$ can lead also to a larger
pre-equilibrium proton emission. Hence finally, due to lowering of Coulomb repulsion 
among the reaction partners, they can interact for a longer time,
favouring the occurrence of dissipative mechanisms \cite{Salvo}.
 
Indeed, in the 52 AMeV \nin case, at $t=80$ fm/c (b = 5 fm),
 a more dissipative neck dynamics is observed in the asy-stiff case.
At b = 8 fm the reaction shows essentially a binary character for both EOS. 
Similar effects are observed for the \nin reaction at 74 AMeV.
In  the \au case, we face a different situation, 
since now the system is neutron-rich, apart from the fact that it is also 
asymmetric in mass and has a larger size. 
For neutron-rich colliding ions the \so choice leads to a little more
dissipative dynamics. 
We can see in figure~\ref{figure12} that at $b=5$~fm, the reaction appears 
quite dissipative and it is
difficult to distinguish the projectile from the target, especially in the
soft case. At $b=10$ fm, the collision is essentially binary. One can see some
particles in between projectile and target regions, mostly
due to pre-equilibrium emission and the two EOS give very similar results. 
However, in general the difference on the reaction path 
between the \so and \st choice appears quite small and one has to explore the behaviour of other 
observables more sensitive to the symmetry energy. 

\subsection{Observables}

\subsubsection{Mass and excitation energy of primary QP/QT}

\begin{figure}[htbp]
\resizebox{0.7\textwidth}{!}{%
\includegraphics{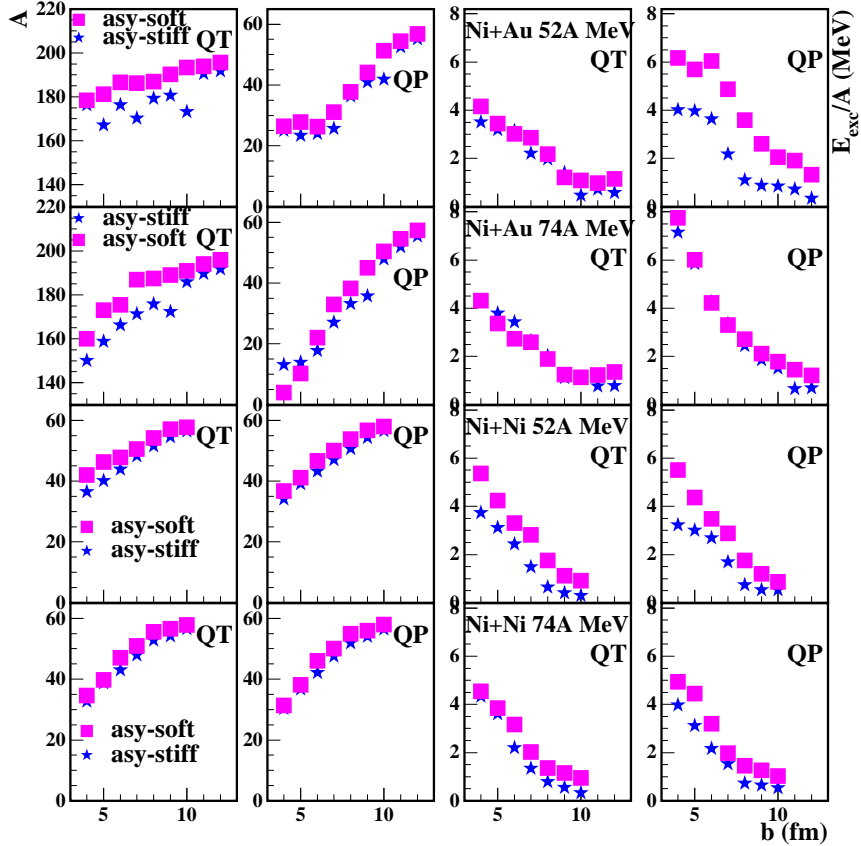}}
\caption{(color online) Masses (columns 1 and 2, left) and
excitation energy per nucleon (columns 3 and 4, right) of the 
quasi- target and  the quasi-projectile as a 
function of the impact parameter, for the two reactions and the two 
energies. The stars correspond to the \st case and the
squares to the \so case.}
\label{figure2}
\end{figure}

We have simulated the reactions until $t=200$~fm/c. The
properties of the two main partners of the collision are considered at 
the time when they re-separate after interaction. We will call this
time $t_{sep}$. At $t_{sep}$, which differs for each impact parameter, QP 
and QT are well defined. A clusterization procedure (in $\vec{r}$ space) was 
used to separate the different products of the reaction~\cite{Bonasera}. 
Due to the mean-field approximation, only heavy
fragments and IMF's can be reconstructed with
this procedure in a reliable way, while the yield of complex particles
is underestimated and the number of free nucleons overestimated.
Thus we can obtain the mass, charge and
excitation energy of the two main partners and all possible fragments. 
The mass and the excitation energy of QT and QP are represented 
in fig.~\ref{figure2} for the two systems, the two energies and the two 
choices of symmetry energy term, as a function of the impact parameter.     

The mass of QT and QP decreases when going towards more central 
collisions. This is due firstly to pre-equilibrium particle emission. 
Moreover, in some cases, an IMF can originate from the overlap region. 
For the \nin system the mass of both QP and QT is little sensitive 
to the choice of the interaction.
On the contrary for the \au system the two equations give
different results for the QT. For both incident energies the mass in the \so case 
is higher than in the \st case. Indeed
with the \st equation the production of an IMF between the two partners 
is more probable, with respect to the re-absorption of the neck region.
This IMF comes essentially  from the target; thus its emission does not affect 
the mass of the QP, whose mass is then independent of the
chosen equation of state. \\   
The excitation energy per nucleon of QT and QP
becomes higher for central collisions in all cases. 
For the symmetric Ni+Ni system, the energies of the two 
partners are equal, as expected. They do not depend on the EOS at 74~AMeV, 
while they are higher with the asy-soft EOS at 52~AMeV, 
due to the less energetic pre-equilibrium emission in this case.
In the Ni+Au system, at 52~AMeV, the excitation energies of the QP
strongly depend on the EOS,  
indicating that more dissipation occurs in the \so case. The effect 
is less evident on the QT side, as expected, since the percentage of
nucleons involved in dissipative mechanisms (in these semi-peripheral
collisions) is less for the system with the largest mass. 
It is interesting to notice also that, at 74~AMeV, the QT keeps
almost the same value of excitation energy as the one obtained  at  52~AMeV,
while only for the QP an increase of excitation energy is observed for the
most central collisions. 
At 74~AMeV the two asy-EOS lead to very similar results.
   
Globally, at the highest incident energy and for both systems, 
probably due to the shorter interaction times, no influence of the EOS 
appears on the excitation energies. Conversely the symmetry term 
does act on the excitation energies at 52~AMeV.

\subsubsection{N/Z ratio of the quasi-projectile}

\begin{figure}[htbp]
\resizebox{0.7\textwidth}{!}{%
\includegraphics{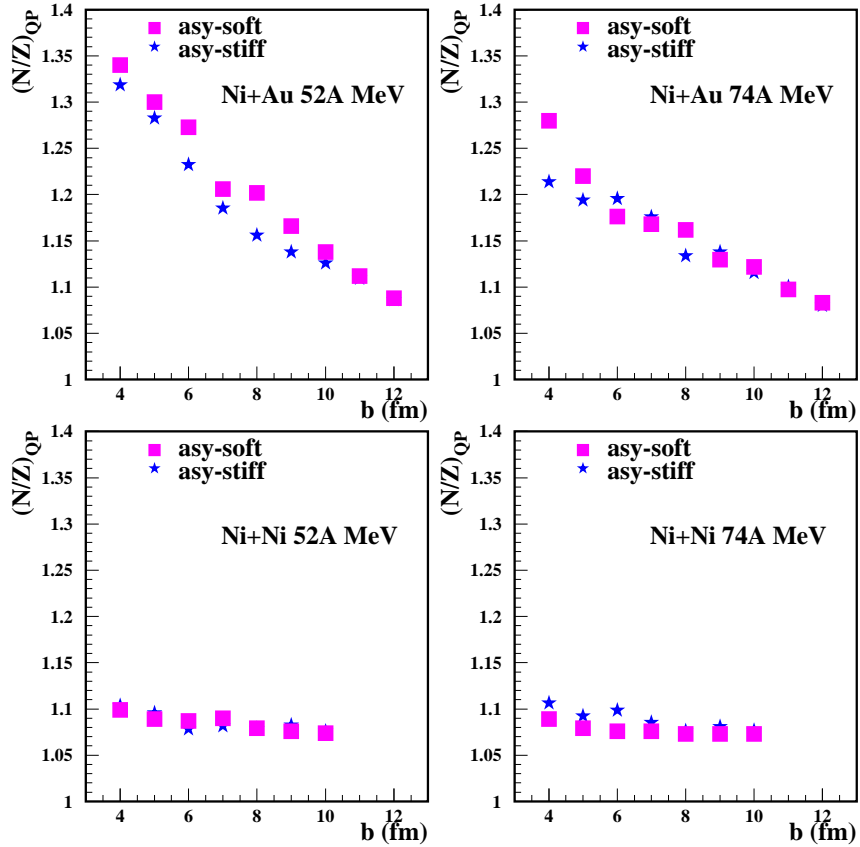}}
\caption{(color online) Isospin ratio of the quasi-projectile as a function 
of the impact parameter, for the two reactions and the two energies. 
The stars correspond to the \st case and the squares to the \so case.}
\label{figure21}
\end{figure}

Let us turn to isospin dependent observables, such as  
the N/Z ratio of the quasi-projectile. 
The isospin ratio of the projectile, $^{58}$Ni, and of the composite Ni+Ni 
system is 1.07 while that of the Au target and of the Ni+Au system is 
respectively 1.49 and 1.38. After collision  the isospin
content of the QP is expected to depend on the target, 
remaining unchanged for the symmetric system, and lying somewhere between 
those of the projectile and of the composite system, depending on the 
interaction and the isospin equilibration times for the Ni+Au system. 
This observable is represented in fig.~\ref{figure21}.

The N/Z ratio increases with the centrality of the collision
for the two systems and the two energies. \\
For the proton-rich \nin system the variation of N/Z with centrality is 
small, and attributed to pre-equilibrium emission. Little dependence on the 
EOS appears at 52~AMeV, while N/Z grows slightly higher at 74~AMeV for the stiff case.
Indeed, with a stiff EOS, more protons are emitted during the pre-equilibrium 
stage. 
This effect increases with the incident energy. On the contrary,
the \so case tends to emit more preequilibrium neutrons leading to a lower
N/Z ratio \cite{lionti}. 
The evolution with centrality is much more pronounced for the neutron-rich
and asymmetric Ni+Au system. In addition to pre-equilibrium effects,  
isospin transport takes place between the two partners of the collision, 
which increases with the violence of the collision. 
N/Z is always higher in the \so  than in the \st case  
(which is less dissipative, as seen above) for the two energies. 
Thus the N/Z diffusion 
appears related to the degree of dissipation reached in 
the system and to the driving force provided by the symmetry term
of the nuclear EOS, that speeds up the isospin equilibration among 
the reaction partners \cite{baranPR,TsangPRL92,Bao-An}.
The largest value reached, at b=4~fm, is lower at 74~AMeV than at 52~AMeV;
this may be attributed to the shorter reaction times, and to the fact that
the collision becomes more transparent. It must be underlined that isospin 
equilibration is nearly reached at the lower energy for the soft EOS, at 
b=4~fm. 
An asy-soft EOS thus favours isospin equilibration between the two partners,
as found also in other recent theoretical investigations
\cite{ShiPRC68,TsangPRL92,Bao-An,Virgil_new}.\\
In conclusion we can say that the effect of the EOS on the quasi-projectile 
N/Z content appears essentially in two
ways: in an asymmetric system (\au case) the effect will be seen 
mostly on the isospin equilibration between the two partners of the collision, 
whereas for a symmetric
system, the effect will react essentially on the pre-equilibrium emission.

\section{Comparison with experimental data}

\subsection{Excitation energy}

\begin{figure}[htbp]
\resizebox{0.7\textwidth}{!}{%
\includegraphics{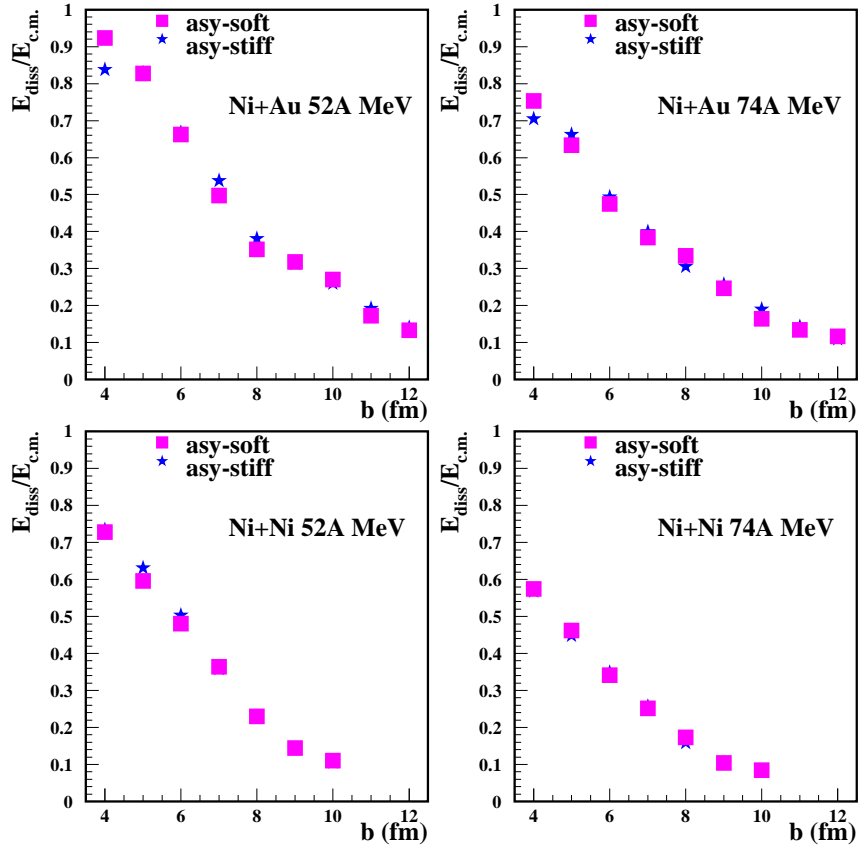}}
\caption{(color online) Correlation between \excm and the impact parameter,
 for the two reactions and the two energies. The stars correspond to the \st 
case and the squares to the \so case.}
\label{figure3}
\end{figure}

In order to compare results of the present model with experimental data
collected with the INDRA detector, the same sorting must be adopted \cite{galichet}. 
The energy dissipated in the reaction was chosen, calculated in the same 
way as in the experimental analysis, where it  is determined from the relative 
velocity between QT and QP. 
The correlation between the calculated dissipated energy normalized to the
centre of mass energy - called \excm
in the following -  and the impact 
parameter is displayed in fig.~\ref{figure3}.   
In all cases the two quantities are strongly correlated, which confirms
that \ex is a good measure of the centrality of the collision.
It should be stressed that a given relative dissipation \excm
corresponds for the different reactions to different impact parameters.
The correlation does not depend much on the employed equation of state, 
although for the \au system a tendency towards more dissipation in 
the \so case starts to be visible below b=5~fm. So, this sorting variable
does not reflect exactly the dissipation energy deposited into the
system, that, as seen in fig.~\ref{figure3}, does depend on the asy-EOS,
especially for the 52~AMeV reactions. This is due to the fact that, 
in the evaluation of \ex, the effects of pre-equilibrium emission are neglected.
  
With the help  of the dissipated energy \ex as a sorting variable, the
behaviour of some observables 
can be followed versus the violence of the collision. For comparisons 
between calculated and experimental data the hot primary quasi-projectiles were
cooled down with the 
proper part of the SIMON code~\cite{simon}.\\

\subsection{Charge of the quasi-projectile}

\begin{figure}[htbp]
\resizebox{0.7\textwidth}{!}{%
\includegraphics{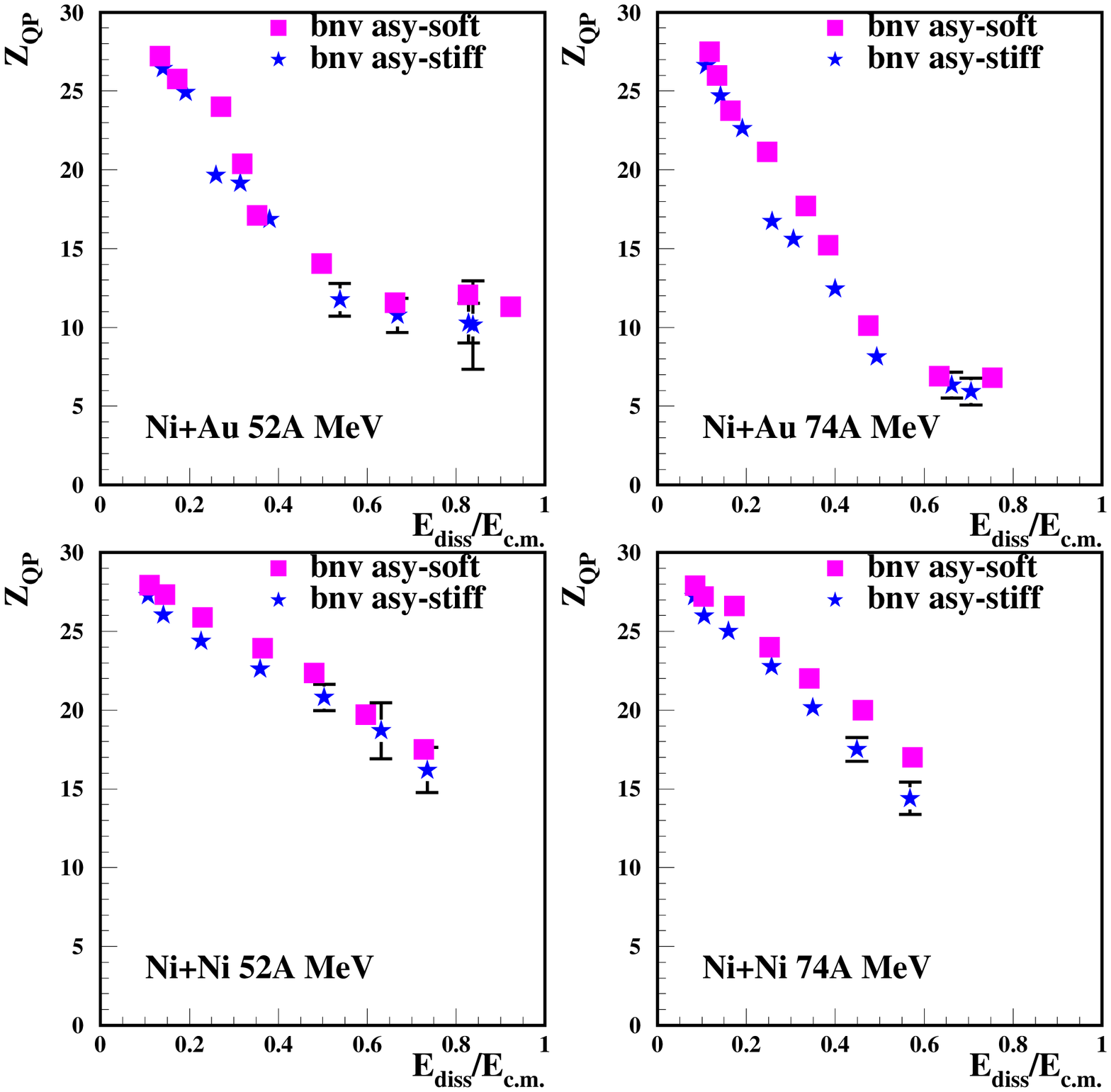}}
\caption{(color online) Calculated charge of the primary quasi-projectile  
vs \excm, for the two reactions and the two energies. The stars 
correspond to the \st case, the squares to the \so case.}
\label{figure4}
\end{figure}

Fig.~\ref{figure4} shows the calculated charge of primary QP 
(defined at $t_{sep}$) as a function of dissipation. 
\begin{figure}[htbp]
\resizebox{0.7\textwidth}{!}{%
\includegraphics{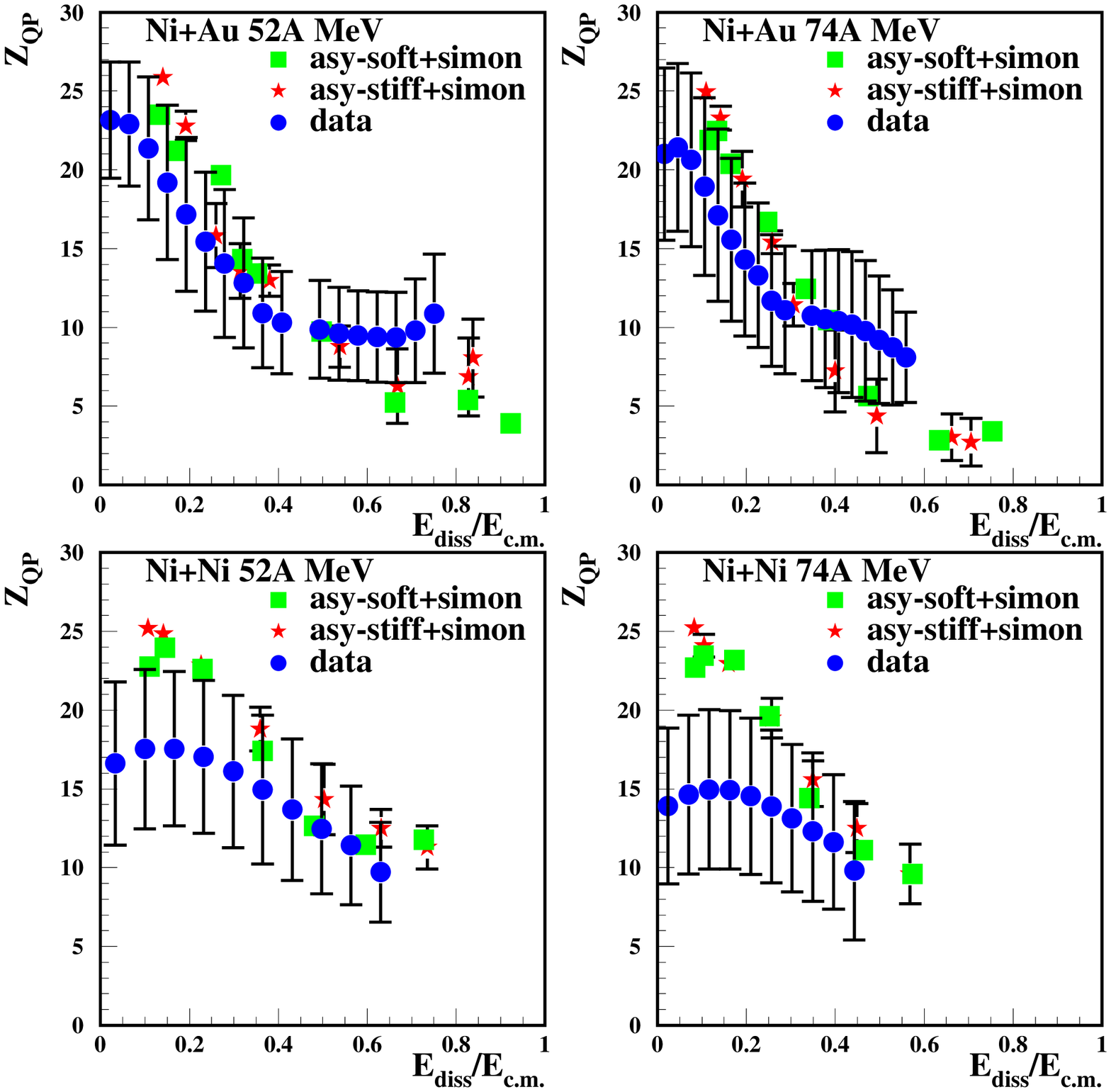}}
\caption{(color online) Charge of the quasi-projectile residue 
vs \excm, for the two reactions and the two energies. The squares
correspond to the \so case+SIMON, the stars to the \st case+SIMON  
and the circles to the experimental data.}
\label{figure41}
\end{figure}
Fragments coming from the neck region were not considered in the evaluation 
of $Z_{QP}$; it was however checked that taking them into account changes
very slightly the QP charge values:
the variation remains within the statistical errors, 
which are represented only
in the \st case in fig.~\ref{figure4} for a better visibility. 
These errors only exceed the size of the symbols for the more dissipative
collisions.\\
As expected, $Z_{QP}$ decreases with the increase of \ex for the two systems and the two energies.
For the \au system the charge 
decreases down to $Z\approx 10$ at 52~AMeV, and to $Z\approx 5$ at 74~AMeV,
 due to a larger pre-equilibrium emission (see sect~II.B.1).  
For the \nin system the charge remains higher ($Z\approx 17$ and 
$Z\approx 15$ for 52 AMeV and 74 AMeV respectively), even at similar 
percentages of dissipation of the available energy.
The two equations of state give rather similar behaviours. However the
soft equation of state gives results that are systematically above the
ones obtained in the stiff case. 
The comparison of the charge of the final QP residue, obtained with the two 
asy-parameterizations, with the experimental values is shown in 
fig.~\ref{figure41}. For experimental data, averages and variances of the
distribution of the heaviest fragment measured by INDRA \cite{galichet} are
displayed. Because of geometry and trigger effects (see \cite{galichet}) 
the more peripheral collisions were strongly rejected for the \nin system, leading 
to an apparent decrease of the charge of the detected QP for the lower 
values of the excitation energy. At higher dissipation, for the two 
systems, data well follow the theoretical trend and charges of final QP 
are within the variance distributions of the data. In all cases 
one can stress that the agreement between calculation and experiment is better for the more 
central collisions. \\ 

\subsection{N/Z ratio as a  function of the dissipation}

\begin{figure}[htbp]
\resizebox{0.7\textwidth}{!}{%
\includegraphics{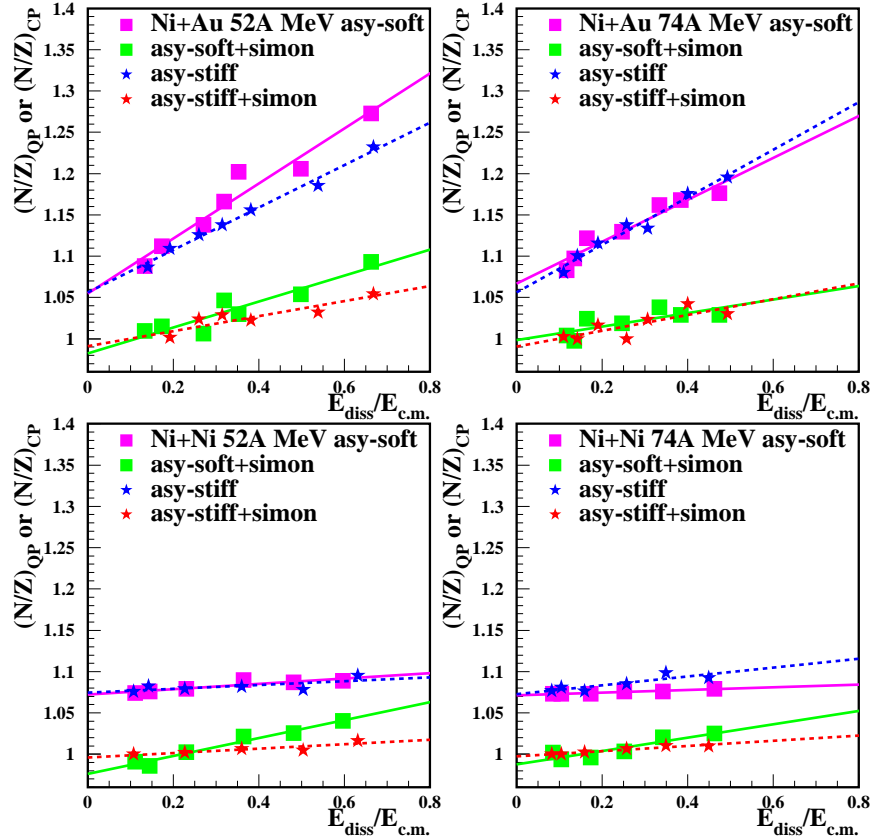}}
\caption{(color online) Isospin ratio of the quasi-projectile vs \excm, 
for the two reactions and the two energies. For the \st calculation, black 
stars and dotted lines display (N/Z)$_{\mathrm{QP}}$ and grey stars and dotted
lines the (N/Z)$_{\mathrm{CP}}$ (BNV calculation followed by SIMON). Same
conventions  for the asy-soft case displayed by squares and full lines. The
lines correspond to linear fits.}
\label{figure5}
\end{figure}

The N/Z ratio of primary QP as a function of
\ex is reported in fig.~\ref{figure5}. The lines represent the results of a linear fit. 
These results, plotted versus b,
 were already discussed in sect.II.B.2; the observed trends are not 
modified when the dissipation is used as a scale for the violence 
of the collision (see also sect. III.A). Particularly, N/Z equilibration is nearly reached in
the Ni+Au system at 52 AMeV, with an asy-soft EOS,  when about 80\% of the
available energy has been dissipated. In the same figure are also plotted
the results concerning the de-excitation step, with the variable called
(N/Z)$_{\mathrm{CP}}$ defined for experimental data. Isotopes 
included in the calculation of the variable are those evaporated by the hot QP.
The values of (N/Z)$_{\mathrm{CP}}$ are always smaller than the N/Z of 
the primary QP. At low dissipation, in all cases, the value starts at 1 instead
of 1.07, due to the dominance of $\alpha$ particles. The evolution of 
(N/Z)$_{\mathrm{CP}}$ with \excm is generally flatter than the (N/Z)$_{\mathrm{QP}}$, 
however the differences between the results of the two parameterizations
are more pronounced for (N/Z)$_{\mathrm{CP}}$, with respect to (N/Z)$_{\mathrm{QP}}$,
 especially at 52 AMeV.  
This is because excitation energies are larger in the \so case and this favours the emission
of neutron-richer particles, thus enhancing the effect due to the larger N/Z value observed
in the \so case for the QP. Only for the reaction Ni+Ni at 74 AMeV, where the N/Z of the
QP is higher in the \st case, the effects due to the de-excitation modify the
initial trend imposed by the dynamical evolution. For the Ni+Au case at 52 A MeV, for instance, the slopes associated with the evolution
of (N/Z)$_{\mathrm{QP}}$ and (N/Z)$_{\mathrm{CP}}$ with  \excm,
obtained with the two parameterizations, differ by 20\% and  40\%, respectively.  
(N/Z)$_{\mathrm{CP}}$ is thus a good witness of isospin transport effects 
and is sensitive to the asy-EOS, though no direct conclusion concerning 
the reach of isospin equilibration among the two reaction partners can be derived 
from this variable. It should be noticed that the difference observed for (N/Z)$_{\mathrm{QP}}$,
between the two parameterizatons, is in agreement with 
recent calculations performed on other systems in the same energy range \cite{Virgil_new}.

\begin{figure}[htbp]
\resizebox{0.7\textwidth}{!}{%
\includegraphics{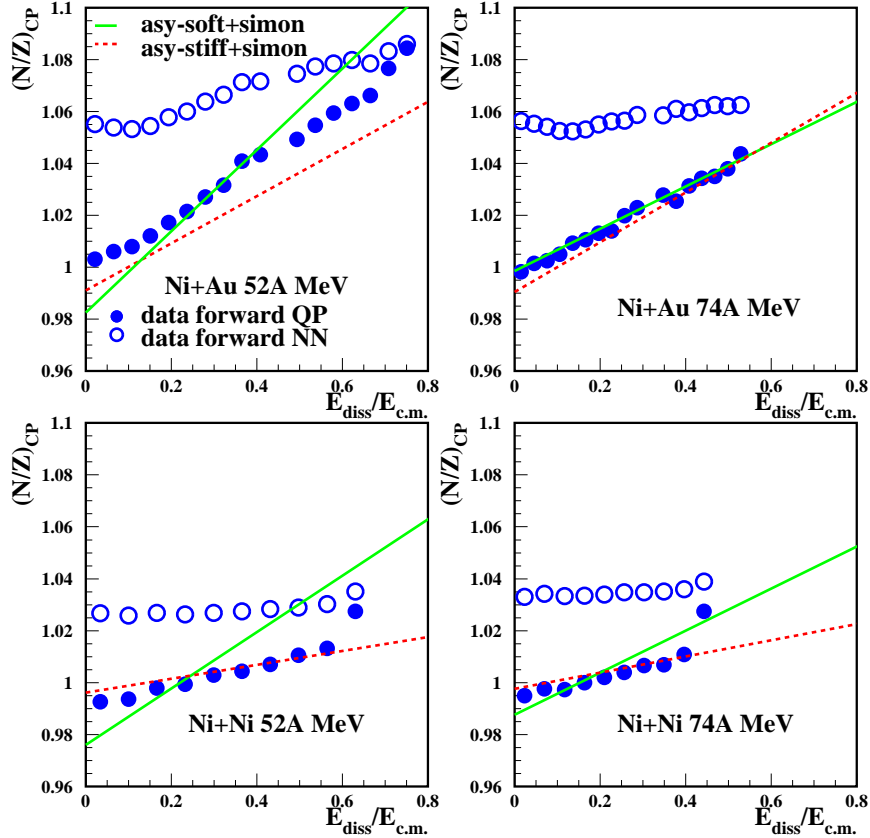}}
\caption{(color online) Isospin ratio of complex particles for Ni
 quasi-projectile vs \excm, for the two reactions and the two energies.
Circles correspond to the experimental data, open for data forward of the 
N-N velocity~\cite{galichet}, full for data forward in the QP frame.
Dotted lines and full lines as in fig.~\ref{figure5}.}
\label{figure51} 
\end{figure}

A comparison of calculated and experimental values of (N/Z)$_{\mathrm{CP}}$ 
is shown in fig.~\ref{figure51}. The results presented in the accompanying 
paper~\cite{galichet} correspond to open circles; they are above the calculated
(N/Z)$_{\mathrm{CP}}$ at low dissipation but values get closer at high dissipation. We remind that these experimental values correspond
to particles forward emitted in the nucleon-nucleon frame, because of the
difficulty to define a QP source. For comparison with the values calculated with
BNV+SIMON, we need a variable more representative of the QP de-excitation
properties. In this aim a second experimental value of (N/Z)$_{\mathrm{CP}}$ was
built with particles forward emitted in the QP frame. The values are remarkably close to the 
BNV+SIMON data, for all systems, particularly for the \st case. They are lower than those built with 
particles emitted forward of the N-N velocity at low dissipation which 
indicates that indeed mid-rapidity emission is more neutron-rich for
the complex particles considered in the variable (N/Z)$_{\mathrm{CP}}$.
Both experimental values (forward N-N and forward QP) present the same trend,
namely no or very little evolution with dissipation for Ni+Ni, and an increase with
the violence of the collision for Ni+Au. For Ni+Au at 52 AMeV, they become equal
at high dissipation. This gives a strong experimental indication that N/Z
equilibration has been reached.

\subsection{Summary of the main findings}

Following the comparison between calculations with two EOS and between calculations and
experimental data, several important points can be stressed.\\
- BNV calculations show a very good correlation between the impact parameter and the variable
\ex calculated as in experiment: \ex appears as a good indicator of the violence of the
collision.\\
- The differences between the two EOS are small, which reflects 
the fact that in the studied collisions nuclear densities keep values 
close to the normal one.\\
- The secondary decay effect is very important and, in almost all cases,
it reduces the sensitivity of the proposed isospin observable 
(N/Z)$_{\mathrm{CP}}$ with \excm. \\
- Calculations support the experimental observation of the significant increase of the
(N/Z) observable for QP in the Ni+Au reactions as a function of the
centrality of the collisions. This appears related to isospin equilibration
between the two reaction partners.\\
- The value of (N/Z)$_{\mathrm{CP}}$ calculated with BNV+SIMON well matches the experimental
data obtained from the forward emitted products in the QP frame for the whole dissipation range
studied. This indicates that the products emitted forward in the QP frame are well 
representative of the QP de-excitation properties.\\
- Globally the asy-stiff case
better matches the experimental data for both systems.
In particular, the behaviour of (N/Z)$_{\mathrm{CP}}$ with
respect to \excm is better reproduced by the \st interaction. A similar
interaction with free nucleon-nucleon cross section was also used in a BUU
transport code to well reproduce isospin diffusion deduced from an
isoscaling analysis for $^{124}$Sn+$^{112}$Sn at 50~AMeV\cite{Bao-An}.\\
- BNV calculations show that isospin equilibration is quasi-reached at the higher dissipation
(impact parameter around 5 fm ) for the Ni+Au system at 52~AMeV. The same conclusion can be
directly deduced experimentally from
the observation that the (N/Z)$_{\mathrm{CP}}$ forward of the 
N-N velocity and forward in the QP frame become equal.\\
- Finally, the isospin equilibration time for reactions in the Fermi energy
domain, as considered here, can be estimated at 130 $\pm$ 10 fm/c;
 this time is the time interval when the 
di-nuclear system remains in interaction, for the most dissipative
binary collisions studied at 52~AMeV.\\

\section{Conclusion}

In this paper we have studied isospin effects in 
semi-peripheral nuclear collisions above the Fermi energy,
with different conditions of mass and charge asymmetry, using different 
equations of state~:~ asy-soft and asy-stiff. 
In this aim, we have compared results obtained on     
quasi-projectiles, in two different reactions with the same projectile:
$^{58}$Ni + $^{58}$Ni and $^{58}$Ni + $^{197}$Au, at incident 
energies of 52 and 74 AMeV. The present analysis is an alternative to
both the isoscaling analysis obtained from an average experimental
impact parameter~\cite{TsangPRL92,Bao-An} and the
pre-equilibrium neutron-to-proton ratios which correspond to events distributed
over an impact parameter range~\cite{Fam06,Bao06,Zha07}. 
Here we study isospin transfer as a function of dissipation or
centrality in collisions  
for two beam energies, looking directly at the average isotopic content
of the emitted light particles.
Simulations show that for the Ni + Ni system, the N/Z of the quasi-projectile
is essentially determined by proton rich pre-equilibrium emission,
and so the N/Z slightly increases with centrality. 
The effect is more pronounced using an \st equation of state.
For the Ni+Au system isospin transport takes place
and the N/Z is larger in the \so case.
Excitation energies are also larger in the~\so case, 
increasing the N/Z of the emitted particles. 
Hence in the Ni+Ni case we observe a kind of compensation 
between the trend imposed by the dynamical evolution and the
secondary decay, while in the Ni+ Au  case, the two effects act in the 
same direction. Finally we find a better overall agreement with
experimental data for the \st case corresponding to a symmetry term linearly
increasing with nuclear density. Moreover more precise information
concerning the isospin equilibration time, as compared to the conclusions of the
experimental joint paper, is obtained. At 52~AMeV for the Ni+Au and the most
dissipative collisions we can infer from the data-model comparison that
isospin equilibration is reached at 130 $\pm$ 10 fm/c. Another very interesting
result comes out from the present study: as far as the N/Z content 
is concerned, the
chemical composition of the quasi-projectile forward emission appears as a very good representation
of the composition of the entire
quasi-projectile source. Such an observation seems to validate a posteriori this
kind of selection frequently used to characterize the properties of
quasi-projectiles.

\end{document}